\documentstyle[prl,aps,amssymb,epsfig,graphicx]{revtex}

\def\be{\begin{equation}}
\def\ee{\end{equation}}
\def\bea{\begin{eqnarray}}
\def\eea{\end{eqnarray}}
\def\eqdef{\stackrel{\mbox{\tiny def}}{=}}     
\def\eqlaw{\stackrel{\mbox{\tiny (law)}}{=}}     

\newcommand{\ket}[1]{|\kern.3ex#1\kern.3ex\rangle}
\newcommand{\bra}[1]{\langle\kern.3ex #1 \kern.3ex|}

\newcommand{\mean}[1]{\left\langle #1 \right\rangle} 

\newcommand{\EXP}[1]{{\mbox{\large e}}^{#1}}         

\def\I{{\rm i}}                  
\def\D{{\rm d}}                  

\begin{document}
\preprint{LPTMS-98/11}
\twocolumn[\hsize\textwidth\columnwidth\hsize\csname@twocolumnfalse\endcsname


\title{Universality of the Wigner time delay distribution \\
       for one-dimensional random potentials}

\author{Christophe Texier$^\dagger$ and Alain Comtet$^\ddagger$}
\address{Laboratoire de Physique Th\'eorique et Mod\`eles Statistiques.\\
         Universit\'e Paris-Sud. B\^at. 100. 91405 Orsay c\'edex. France.}
\date{\today}
\maketitle


\begin{abstract}
We show that the distribution of the time delay for one-dimensional random
potentials is universal in the high energy or weak disorder limit.
Our analytical results are in excellent agreement with extensive numerical
simulations carried out on samples whose sizes are large compared to the
localisation length (localised regime). The case of small samples is also
discussed (ballistic regime). 
We provide a physical argument which explains in a quantitative way the
origin of the exponential divergence of the moments.
The occurence of a log-normal tail for finite size systems is analysed. 
Finally, we present exact results in the low energy limit which clearly show
a departure from the universal behaviour.
\end{abstract}

\twocolumn
\vskip.5pc]


The problem of quantum scattering by chaotic or disordered systems is
encountered in many fields ranging from atomic or molecular physics  as
well as in the scattering of electromagnetic microwaves. Some properties of
the scattering process are well captured through the concept of time delay.
This quantity, which goes back to Eisenbud and Wigner \cite{Eis48Wig55},
is related to the time spent in the interaction region by a wave packet of
energy peaked at $E$. It can be expressed in terms of the derivative of the
$S$ matrix with respect to the energy. In the context of chaotic scattering
the approach based on random matrix theory (RMT) provides a statistical
description of the time delays. This problem was first studied by a
supersymmetric approach \cite{FyoSom96a} and in \cite{GopMelBut96} by using a
statistical analysis. This latter work provides a derivation for
the one channel case for the different universality
classes. Recently it served as a starting point for \cite{BroFraBee97}
where the $N$ channel distribution is shown to be given by the Laguerre
ensemble of RMT.
In spite of its success, such a description by RMT is not entirely
satisfactory, in particular it does not apply to strictly one-dimensional
systems \cite{FyoMir94} for which strong localisation effects occur.
Furthermore it does not shed much light on the physical mechanisms which are
responsible for the universal distribution. In this work
we explore another approach by considering the scattering by a
one-dimensional random potential. In this case the existence of universal
distributions was first conjectured in \cite{ComTex97} on the basis of a
comparative study of two different models. This was further supported by
\cite{JosJay97} where the random potential is still of a different kind.


The purpose of this letter is to present a new derivation that accounts
for the universality and also to provide a physical picture that explains
the origin of the algebraic tail of the distribution in terms of resonances. 
Further details will be given elsewhere \cite{ComTex99}. To begin
with, let us briefly recall the model. We consider the Schr\"odinger
equation on the half line $x\geqslant0$:
\be
-\frac{\D^2}{\D x^2} \psi_k(x) + V(x) \psi_k(x) = k^2 \psi_k(x)
.\ee
We assume that $V(x)$ has its support on the interval $[0,L]$ and 
impose the Dirichlet boundary condition $\psi_k(0)=0$.
Therefore, for $x\geqslant L$ stationary scattering states of the form
\be\label{psik}
\psi_k(x)=\frac12\left(\EXP{-\I k(x-L)}+\EXP{\I k(x-L)+\I\delta(k)}\right)
\ee
represent the superposition of an incoming and a reflected plane wave.
Since there is only backward scattering, the reflection coefficient
$\EXP{\I\delta(k)}$ is of unit modulus and the Wigner time delay takes the
form $\tau(k)\eqdef\frac{1}{2k}\frac{\D\delta(k)}{\D k}$. Such a model with
a random potential can be viewed as a model of a disordered sample connected
to an infinite lead.  Instead of using the invariant
embedding method as in \cite{JayVijKum89,Hei90} or stochastic differential
equations \cite{FarTsa94}, our starting point 
is to relate the time delay to the wave function inside the sample. 
This may be achieved by using the identity
\be
\frac{\D}{\D x}\left(
 \frac{\D \psi^*}{\D x}\frac{\D \psi}{\D E} - \psi^* \frac{\D^2 \psi}{\D x\D E}
\right)=|\psi|^2
.\ee
By integration over $[0,L]$ one gets the so-called Smith formula \cite{Smi60} 
\be\label{refonc4}
\tau(k)=\frac{2}{k}\int_0^L \D{x}\,|\psi_k(x)|^2-\frac{1}{2k^2}\sin\delta(k)
.\ee
It expresses the time delay as the sum of a dwell time \cite{But83} and a
term that can be neglected in the high energy limit.
Inside the sample, the wave function and its derivative may be written in
the form $\psi_k(x)={\cal N}\sin\theta(x)\EXP{\xi(x)}$
and $\psi'_k(x)=k{\cal N}\cos\theta(x)\EXP{\xi(x)}$. The normalisation factor
$|{\cal N}|=\EXP{-\xi(L)}$ is fixed by matching the wave function at $x=L$
with the scattering states (\ref{psik}).
We now consider the case where $V(x)$ is a random potential.
In this case the growth or decay of the envelope $\EXP{\xi(x)}$ of the wave
function is measured by the Lyapunov exponent 
$\gamma$ (inverse localisation length $\lambda=\gamma^{-1}$). 
In the high energy limit, the envelope is a slow variable,  while
the phase $\theta(x)$ presents rapid oscillations on a scale $k^{-1}$.
Therefore, in the high energy limit one can integrate over the fast variable
in (\ref{refonc4}) and get:
\be\label{THErepfonc}
\tau(k)=\frac{1}{k}\int_0^L \D{x}\,\EXP{2(\xi(x)-\xi(L))}
.\ee
This representation of the time delay holds for any realization of
the disordered potential. It therefore captures all the statistical
properties of $\tau(k)$ once the distribution of $\xi(x)$ is known. Denoting by
$x_c$ the correlation length of $V(x)$ and assuming that $x_c$ and $k^{-1}$
are the smallest length scales of the system, then it was proven in
\cite{AntPasSly81} that  the variable $\xi(x)$ is a
Brownian motion of the form: $\xi(x)=\gamma x +\sqrt\gamma W(x)$ where
$W(x)$ is a normalised Wiener process ($\mean{W(x)}=0$,
$\mean{W(x)W(x')}=\min(x,x')$). Thus the Lyapunov exponent $\gamma$ controls
both the drift and the fluctuations. Using the scaling
properties of the Brownian motion then gives the following identity in
law: 
\be\label{rf}
\tau(k)\eqlaw\frac{1}{k\gamma}\int_0^{\gamma L} \D{u}\,\EXP{-2u+2W(u)}
.\ee
This representation of the time delay as an exponential functional of
the Brownian motion \cite{Yor92,MonCom94,ComMonYor98} allows to derive a
number of interesting results:\\
({\em i})~existence of a limit distribution ($\tau$ fixed, $L\to\infty$)
with an algebraic tail \cite{Pastur}:
\be\label{Ps}
P(\tau)=\frac{\lambda}{2k\tau^2}
\EXP{-\frac{\lambda}{2k\tau}}
.\ee
({\em ii})~Linear divergence of the first moment and exponential divergence 
of the higher moments \cite{MonCom94}:
\bea
\mean{\tau(k)}   
    &=& \frac{L}{k} \\ 
\mean{\tau(k)^n} 
    &=& \bigg\{ \sum_{m=2}^n(-1)^{n-m} C_n^m 
	\frac{(m-2)!\,(2m-1)}{(n+m-1)!}\EXP{2m(m-1)L/\lambda} \nonumber \\
    & & \hspace{0.8cm} + 
	\frac{(-1)^{n+1}}{n!}\left( 2n\frac{L}{\lambda}+n-1 \right)
        \bigg\}\left(\frac{\lambda}{2k}\right)^n
.\eea
In \cite{ComTex97} we have shown that these results hold for two different
models of random potential in the localised regime ($L\gg\lambda$).

\begin{figure}[!h]
\begin{center}
\includegraphics[scale=0.3,angle=-90]{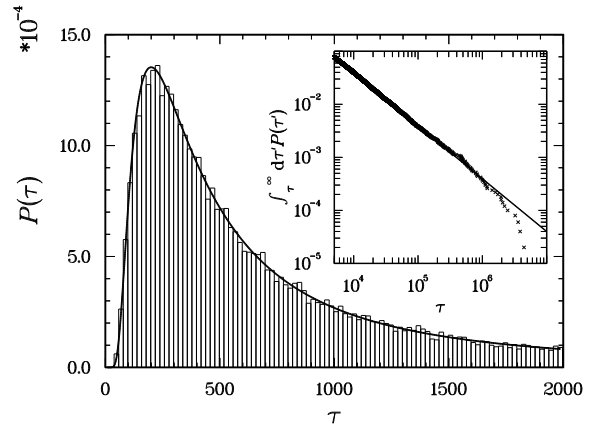}
\caption{Time delay distribution in the localised regime $L\gg\lambda$. 
         Comparison between the numerical calculation and expression
	 (\ref{Ps}). In the inset: tail of the integrated distribution,
	 numerical and analytical.}\label{idtid0} 
\end{center}
\end{figure}

In order to test the analytical results in the above mentioned regime 
it is convenient to choose a model suitable for numerical simulations. For
this purpose we 
have considered the case where the random potential is given by a sum of delta
functions of the same weight $v$, randomly dropped on the half line with an
average density $\rho$ (the so-called Frish and Lloyd model)\footnote{
This model coincides with the Gaussian model (Halperin model)
considered in \cite{ComTex97} in the limit of a high density of impurities 
($v\ll k\ll\rho$).}.  
The equations
that describe the evolution of the phase take a discrete form which can
be implemented conveniently in a numerical simulation. The distribution of
the time delay that we have obtained numerically in this way is in
perfect agreement with equation (\ref{Ps}) as soon as the high energy regime is
reached. For example we compare on figure \ref{idtid0} the analytical
expression (\ref{Ps}) with the corresponding numerical result for a regime 
$\rho\ll v\ll k$. The simulation was carried out for a system with $10^5$
impurities 
of weight $v=1$, distributed with an averaged density $\rho=0.1$. 
The energy considered corresponds to $k=10$ which is related to a
localisation length $\lambda=\frac{8k^2}{\rho v^2}=8000$. The ratio
$L/\lambda=125$ is sufficiently large for the limit distribution to be
reached. The numerical calculation is based on a statistics of $50000$
values. It shows that the algebraic tail is well reproduced
by (\ref{Ps}) for $2000$ times the typical value 
$\tau_{\mbox{\tiny typ}}=200$. Let us stress that
there is no adjustable parameter to fit the numerics. The only 
parameter entering in the analytical expressions is the localisation length
which is known for each kind of disorder.

The derivation of the statistical properties of $\tau$ given above
allows one to understand the universality of the result but on
the other hand does not shed much light on the physical mechanisms which are
responsible for the occurence of an algebraic tail. In the following we
propose a physical picture based on the existence of resonances that explains
the leading exponential behaviour of the moments. The starting point is to
realize that the reflection of the incident wave on the random potential can
in fact be viewed as a resonance tunneling process
\cite{FriFroSchSul73,Azb83}. Indeed there exists a representation of
the time delay as a superposition of resonances of energy $E_\alpha$ 
and width $\Gamma_\alpha$ in the form \cite{FyoSom96b}
\be\label{repFyo}
\tau(E)=2\sum_\alpha\frac{\Gamma_\alpha/2}{(E-E_\alpha)^2+\Gamma_\alpha^2/4}
.\ee
Obviously the dominant contribution
$\tau\simeq\frac{4}{\Gamma_\alpha}$ is achieved when $E$ is in a window of
width $\Gamma_\alpha$ centered at $E_\alpha$, and this will
occur with 
probability $\frac{\Gamma}{\Delta}$ where $\Delta$ is the mean level
spacing. In order to estimate the width we may assume that a discrete level
$E_\alpha$ localised at $x_0$ will be broadened by its coupling to the
continuum 
of states through the end point $x=L$. We may therefore set
$\Gamma\sim\EXP{-2\gamma_L(L-x_0)}$ where $\gamma_L$ is the Lyapunov
exponent in the finite system. Assuming that $x_0$ is uniformily distributed
on $[0,L]$ and decorrelated from $\gamma_L$ one obtains the estimate:
\be
\mean{\tau^n} 
\sim \int_0^L\frac{\D x_0}{L}\int\D\gamma_L\,
     p(\gamma_L)\frac{\Gamma}{\Delta}\frac{1}{\Gamma^n}
.\ee
Since $\xi(L)=L\gamma_L$ defined previously is a Gaussian process, the
distribution of the finite size Lyapunov exponent $\gamma_L$ is
\cite{AntPasSly81}
\be
p(\gamma_L)=\sqrt{\frac{L}{2\pi\gamma}}
\EXP{-\frac{L}{2\gamma}(\gamma_L-\gamma)^2}
.\ee
One finally obtains 
\be
\mean{\tau^n} \sim  \EXP{2n(n-1)L/\lambda}
.\ee
A more refined derivation \cite{ComTex99} allows to recover the gross
behaviour of the pre-exponential factor.
This demonstrates that this
particular behaviour of the moments has his origin both in the
exponentially small widths of the resonances and in the fluctuations of the
Lyapunov exponent for the finite size sample.


This physical picture shows that the  problem of the time delay distribution
related to the question of quantum relaxation in
disordered systems. In this respect, the fact that $P(\tau)$ is a broad
distribution is rather puzzling. Indeed it was recently argued that at
large time, the probability distribution of various quantities, such as the
conductance of one-dimensional disordered sample, exhibits a log-normal tail
\cite{MuzKhm97}. 
In order to clarify this point, instead of considering as before the regime
$\tau$ fixed $L\to\infty$ which leads to (\ref{Ps}), we have studied for
fixed $L$ the tail of the distribution in the limit $\tau\to\infty$. 
In order to extract the asymptotic
behaviour, it is  convenient to consider the characteristic function
$\phi(p,L)=\int_0^\infty\D\tau\,\EXP{-2kp\tau}P(\tau;L)$ given in
\cite{MonCom94}. If the conjugated variable $p$ is chosen in a range
$\gamma\EXP{-\gamma L}\ll p\ll\gamma\EXP{-\sqrt{\gamma L}}$,
the characteristic function exhibits the following behaviour
\be
\phi(p,L)\simeq1-\frac{2\sqrt\pi\EXP{-\frac{\gamma L}{2}}}{(2\gamma L)^{3/2}}
\ln{\gamma/p}\left[1+O\left(\frac{\ln{\gamma/p}}{\gamma L}\right)\right]
\EXP{-\frac{\ln^2{\gamma/p}}{8\gamma L}}
,\ee
which suggests the existence of a log-normal tail for the distribution
\be
P(\tau;L) \sim
\exp{-\frac{1}{8\gamma L}\ln^2(2k\gamma\tau)}
\ee
in the range $\EXP{\gamma L}\gg2k\gamma\tau\gg\EXP{\sqrt{\gamma L}}$.
Although we were not able to derive the behaviour of the distribution
when $\EXP{\gamma L}\ll2k\gamma\tau$, the fact that the most divergent part
of the moments grow like $\EXP{2n^2L/\lambda}$ suggests that the
distribution is still log-normal.

As an aside remark let us mention that a log-normal distribution of the time
delay also occurs in the study of the random mass Dirac model at the critical
point $E=0$ \cite{SteCheFabGog98}. There, the authors provide a
representation of $\tau$ which is similar to equation (\ref{rf}),
except that the drift term is absent. This problem can also be analysed by
using the approach given in \cite{MonCom94,ComMonYor98}.


\begin{figure}[!h]
\begin{center}
\includegraphics[scale=0.275,angle=-90]{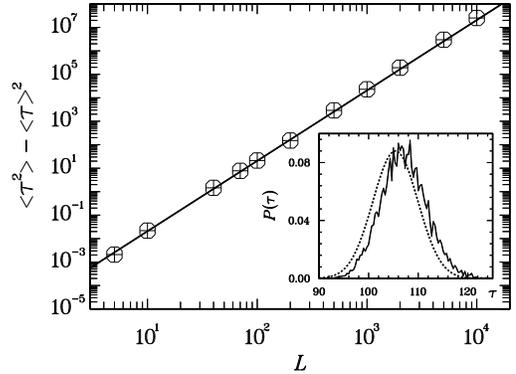}
\caption{Second cumulant of the time delay in the ballistic regime
         $L\ll\lambda$. Comparison between numerical results and the
	 analytical result 
	 $\mean{\tau^2}-\mean{\tau}^2=\frac{\rho v^2}{6(k^2-\rho v)^2}L^3$.
	 the parameters are $v=0.001$, $\rho=100$ and $k=1$.
	 In the inset: time delay distribution for $L=100$ ($10^4$
	 impurities). 
	 }\label{tau2m}
\end{center}
\end{figure}

At this stage we have only considered a localised regime for which the size
of the system is large compared to the localisation  length $L\gg\lambda$.
Another interesting case  is the ballistic one
characterised by $L\ll\lambda$. In this situation, the dimensionless
variable $\gamma L$ which arises in (\ref{rf}) is small compared to 1 and
the argument of the exponential typically remains 
small compared to 1, which allows one to expand the exponential. The
resulting expression for the time delay is
given by a linear functional of a Gaussian quantity
and has itself Gaussian fluctuations characterised by a first moment
$\mean{\tau}=\frac{L}{k}$ and a second cumulant
$\mean{\tau^2}-\mean{\tau}^2\simeq\frac{4\gamma}{3k^2}L^3$.
We have checked numerically these results with the delta impurity model. We
have considered a regime where it reproduces the high energy features of
the Gaussian model: $v\ll k\ll\rho$ and $(k^2-\rho v)\gg(\rho v^2)^{2/3}$.
In this regime one has to take into account the averaged value of the
disorder $\rho v$ and replace $k$ by $\sqrt{k^2-\rho v}$ in all previous
expressions: the localisation length is thus given by
$\lambda=\frac{8(k^2-\rho v)}{\rho v^2}$ and the moments of $\tau$ now read:
$\mean{\tau}=\frac{L}{\sqrt{k^2-\rho v}}$ and
$\mean{\tau^2}-\mean{\tau}^2\simeq\frac{\rho v^2}{6(k^2-\rho v)^2}L^3$.
In figure \ref{tau2m} we compare the numerical result to the Gaussian
distribution where the parameters are given by the previous expressions. 
The calculation is performed for a ratio $L/\lambda\simeq1.4\ 10^{-3}$.
$10000$ values of $\tau$ were calculated.


\begin{figure}[!h]
\begin{center}
\includegraphics[scale=0.275,angle=-90]{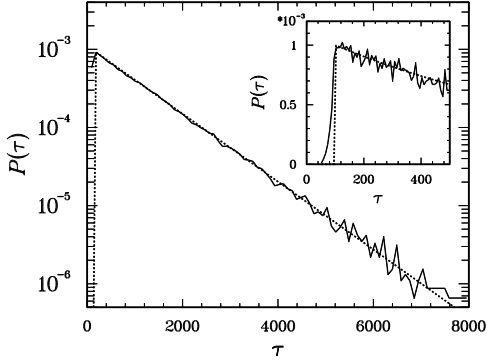}
\caption{Time delay distribution in the localised regime 
         $L\gg\lambda$ at low energy.}\label{idtid1}
\end{center}
\end{figure}

Let us close the paper with some remarks on the low-energy regime.
In this case one is more sensitive to the
precise nature of the disorder, therefore one can guess that universality
must break down. The distribution of the time delay will now depend on 
the nature of disorder. As an illustration let us consider the delta
impurity model; we predict 
\cite{ComTex99}, in the low density regime $k\ll\rho\ll v$, an
exponential tail for the distribution: 
\be\label{dale}
P(\tau)\simeq k\rho\,{\rm Y}(\tau-\frac{1}{kv})\,\EXP{-k\rho(\tau-\frac{1}{kv})}
,\ee
where ${\rm Y}(x)$ denotes the Heaviside function. We have checked that
this expression is in very good agreement with  the numerical results.
The numerical computation was performed for $v=1$, $\rho=0.1$, $k=0.01$ for
$1000$ impurities. The resulting distribution presented on figure
\ref{idtid1} is based on $50000$ data sets.
The behaviour at the origin is more subtle than the one given above,
nevertheless (\ref{dale}) gives the correct scale on which the distribution
vanishes at the origin. 
Another indication that universality breaks down is the fact that
within this same model, the  low energy regime with a high density of
impurities $k\ll v\ll\rho$ leads to different distributions, though still
characterised by an exponential tail.

\bigskip

\noindent
{\bf Acknowlegments}

We thank David Dean, Yan Fyodorov, Spyros Evangelou, Alexander O. Gogolin, 
Leonid Pastur and Marc Yor for interesting discussions. 
We thank Marie-Th\'er\`ese Commault for the realization of figures.

\medskip

\noindent
$^\ddagger$ e-mail: comtet@ipno.in2p3.fr\\
$^\dagger$ e-mail: texier@ipno.in2p3.fr


\end{document}